\documentstyle[aps,epsf]{revtex}
\begin{document}
\draft
%\preprint{}
\title{Explaining critical angular velocities of the vortex formation
in a stirred Bose-Einstein condensate.}
\author{A.~A.~Kozhevnikov \footnote{Electronic address: kozhev@math.nsc.ru}}
\address{Laboratory of Theoretical Physics, \\
Sobolev Institute for Mathematics \\
630090, Novosibirsk-90, Russia}
\date{\today}
\maketitle
\begin{abstract}
The problem of explanation of the critical angular velocity
$\Omega_c$ when the  formation of a vortex in the stirred
Bose-Einstein condensate becomes energetically possible, is
considered in the framework of the variational approach. The
origin of smallness of the calculated $\Omega_c$ in comparison
with the measured values which takes  place for  pure quantum
state with the unit angular momentum per condensed particle,
 is uncovered. The agreement with the measured $\Omega_c$ is
achieved upon admitting a small admixture of the zero angular
momentum state in the wave function of the one-vortex quantum
state prepared after stirring. The portion of this admixture
amounts to 10$ - 13 \%$ of the total condensed atoms. Possible
test of this hypothesis is proposed.
\end{abstract}
%\narrowtext

Bose-Einstein condensate (BEC) discovered in the trapped clouds of
alkali atoms \cite{becexp} are proved to be an excellent test site
of fundamental concepts of quantum physics of the systems
consisting of macroscopically large number of particles
\cite{rmp}. One of the most intriguing features of such systems is
the quantized vortex. Recently, the formation of such vortices in
the trapped BEC was observed in two different situations. First is
the two-component BEC \cite{vortexp1}, where the vortex state is
created via interconversion between two hyperfine states. The
second one is realized by the stirring of BEC with the toggled
laser beam. This toggling beam creates a small axial asymmetry of
the trap potential which is rotated slowly with the angular
velocity $\Omega$. It was found that for $\Omega$ exceeding a
definite critical value $\Omega_c$, the images of the space
distribution of BEC atoms after the ballistic expansion reveals a
visible signature of the vortex \cite{vortexp2}, and even the
lattice of the vortex array was observed in some situiations
\cite{vortexp2}. The method of Ref.~\cite{vortexp2} directly
corresponds to the classical experiment with the rotating bucket
\cite{bucket}. As was pointed out in Ref.~\cite{vortexp2}, the
measured critical angular velocity of the formation of a single
vortex in the stirred BEC appears to be notably larger than that
predicted theoretically \cite{lundh}.

Recently, the efforts aimed at explaining  larger critical angular
velocity were undertaken in Ref.~\cite{effort}. The purpose of the
present note is to propose another explanation of the critical
angular velocity $\Omega_c$ of the stirring of BEC when the
formation of the single vortex  becomes energetically possible. To
this end the energy of different configurations of the BEC atoms
is calculated. The reason of smallness  of the calculated
$\Omega_c$ as compared to the measured one is revealed. It is
shown that the agreement with the measurements can be achieved by
assuming the admixture of the vortex-free state with zero angular
momentum, in the wave function of the final state prepared after
the stirring and containing the visible vortex.

As is known \cite{rmp}, all basic properties of BEC in diluted
gases of alkali metals are described by Gross-Pitaevskii (GP)
equation \cite{gp} which has the form of Schr\"{o}dinger equation
added with the nonlinear term arising due to the short range
interaction characterized by the single parameter - the scattering
length. Since it is  the equilibrium energy of the BEC gas that is
of the main concern here, the GP energy functional
\begin{equation}
E=\int
d^3x\left\{\frac{\hbar^2}{2m}\left|\bbox{\nabla}\psi\right|^2+\frac{m}{2}
\left(\omega^2_\bot
r^2_\bot+\omega^2_zz^2\right)|\psi|^2+\frac{2\pi a\hbar^2}
{m}|\psi|^4\right\}, \label{en}
\end{equation}
is used instead of GP equation. In the above equation,
$r^2_\bot=x^2+y^2$, $m$ is the mass of an atom, $\omega_\bot$,
$\omega_z$ are, respectively, the transverse and longitudinal
frequencies of the oscillator-like potential modeling the axially
symmetric trap, $a$ is the scattering length. Also, $\psi$ is the
condensate wave function normalized according to the condition
\begin{equation}
N=\int d^3x|\psi|^2, \label{norm} \end{equation}$N$ is the number
of condensed atoms. The trap parameters $N$, $\omega_\bot$, and
$\omega_z$ are specified as follows. The first set referred below
as the set A is \cite{vortexp2}
\begin{equation}
N=(1.4\pm0.5)\times10 ^5\mbox{,
}\frac{\omega_\bot}{2\pi}=219\mbox{ Hz,
}\frac{\omega_z}{2\pi}=11.7\mbox{ Hz}, \label{seta}
\end{equation}
and the corresponding critical angular velocity of stirring is
$\Omega_c/2\pi=152$ Hz \cite{vortexp2}. The second set referred
below as the set B is \cite{vortexp3}
\begin{equation}
N=(3.7\pm1.1)\times10 ^5\mbox{,
}\frac{\omega_\bot}{2\pi}=171\mbox{ Hz,
}\frac{\omega_z}{2\pi}=10.3\mbox{ Hz}, \label{setb}
\end{equation}
and the corresponding critical angular velocity of stirring is
$\Omega_c/2\pi=115$ Hz \cite{vortexp3}.

Since the accuracy of determination of the number of condensed
atoms is about 30 $\%$, and the energy of BEC is scaled as
$E\propto N^{2/5}$ [see Ref.~\cite{rmp} and Eqs.~(\ref{gamz}),
(\ref{etf}) below], it is unnecessary to bother to calculate $E$
with the accuracy better than  10$\%$. So, one may hope that the
variational calculation \cite{var} of energy instead of full
numerical solution of GP equation will be sufficient. As will soon
become clear, it is convenient to take the trial wave function in
the form
\begin{equation}
\psi_\kappa(r_\bot,\phi,z)=\left(\frac{N}{\pi^{3/2}R^2_\kappa
z_\kappa}\right)^{1/2}\left(\sqrt{1-\kappa}+\sqrt{\kappa}\frac{r_\bot}{R_\kappa}
e^{i\phi}\right)\exp\left(-\frac{r^2_\bot}{2R^2_\kappa}-\frac{z^2}{2z^2_\kappa}\right),
\label{psi}
\end{equation}
where $0\leq\kappa\leq1$;  $R_\kappa$ and $z_\kappa$ are
variational parameters. Notice that $\kappa=0\mbox{, }1$
corresponds to, respectively, pure  vortex-free state and the
state with the singly quantized vortex placed at the center of the
trap, while intermediate values of $\kappa$ correspond to
arbitrary mixture of the above states. Introducing the
dimensionless parameters $\rho_\kappa$ and $z_\kappa$ according to
the relations
$R_\kappa=\left(\hbar/m\omega_\bot\right)^{1/2}\rho_\kappa$ and
$z_\kappa=\left(\hbar/m\omega_z\right)^{1/2}\zeta_\kappa$, one can
find from Eqs. (\ref{en}) and (\ref{psi}) the energy per condensed
atom:
\begin{equation}
\frac{E(\kappa)}{N}=
\frac{\hbar\omega_\bot}{2}\left(\frac{1}{\rho^2_\kappa}+\rho^2_\kappa\right)(1+\kappa)+
\frac{\hbar\omega_z}{4}\left(\frac{1}{\zeta^2_\kappa}+\zeta^2_\kappa\right)+
\frac{\hbar\omega_\bot\gamma_z}{\rho^2_\kappa\zeta_\kappa}\left(1-\frac{\kappa^2}{2}\right),
\label{e1}
\end{equation}
where
\begin{equation}
\gamma_z=aN\left(\frac{m\omega_z}{2\pi\hbar}\right)^{1/2}.
\label{gamz}
\end{equation}
The mean value of  angular momentum in the quantum state with the
wave function Eq.~(\ref{psi}) is
\begin{equation}
\langle L_z\rangle=\int
d^3x\psi^\ast_\kappa\left(-i\hbar\frac{\partial}{\partial\phi}\right)\psi_\kappa=\hbar
N\kappa. \label{lz}
\end{equation}
As is known \cite{rmp,lif}, the condition of thermodynamic
possibility of the vortex formation in the system rotated at the
angular velocity $\Omega$ can be formulated as $\Delta E-\langle
L_z\rangle\Omega<0$, where $\Delta E$ is the energy difference
between the states with vortex and without it, so that the
critical angular velocity is defined as $\Omega_c=\Delta E/\langle
L_z\rangle$. Let us evaluate $\Omega_c$ in the framework of
variational approach.

The values of variational parameters $\rho_\kappa$ and
$\zeta_\kappa$ can be found from the condition of the minimum of
the BEC energy Eq.~(\ref{e1}) which is reduced to the  following
equations:
\begin{eqnarray}
\rho^2_\kappa-\rho^{-2}_\kappa&=&\frac{2\gamma_z(1-\kappa^2/2)}
{\rho^2_\kappa\zeta_\kappa(1+\kappa)}, \nonumber\\
\zeta^2_\kappa-\zeta^{-2}_\kappa&=&\frac{2\omega_\bot\gamma_z(1-\kappa^2/2)}
{\omega_z\rho^2_\kappa\zeta_\kappa}. \label{varpar}
\end{eqnarray}
First, let us consider the problem of the critical angular
velocity in the approximation when the kinetic energy of both the
transverse and longitudinal motion can be neglected. This is the
Thomas-Fermi (TF) limit \cite{rmp}.  The solution of
Eq.~(\ref{varpar}) in this limit looks as
\begin{eqnarray}
\rho^{\rm TF}
_\kappa&=&\left[\frac{4\gamma_z^2\omega_z(1-\kappa^2/2)^2}
{(1+\kappa)^3\omega_\bot}\right]^{1/10} \nonumber\\ \zeta^{\rm TF}
_\kappa&=&\rho^{\rm TF}
_\kappa\left[\frac{\omega_\bot}{\omega_z}(1+\kappa)\right]^{1/2}.
\label{partf}
\end{eqnarray}
Direct numerical evaluation shows that at, say, $\kappa=0$ the
parameters $\rho^{\rm TF}_0=2.16$ (2.63) and $\zeta_0^{\rm
TF}=9.36$ (10.7) evaluated in this limit for the set of parameters
Eq.~(\ref{seta}) [respectively, (\ref{setb})],  coincide within
the accuracy 1$\%$ with those found from Eq.~(\ref{varpar}).
Hereafter, when doing the specific numerical evaluations, we take
the scattering length $a=5.77$ nm for $^{87}$Rb atoms \cite{a},
and the above two sets of BEC parameters from Eqs.~(\ref{seta})
and (\ref{setb}). The corresponding energy in TF limit is found to
be
\begin{equation}
\frac{E^{\rm
TF}(\kappa)}{N}=\frac{5}{4}\hbar\omega_\bot\left(\frac{4\gamma^2_z\omega_z}
{\omega_\bot}\right)^{1/5}
\left[(1+\kappa)(1-\kappa^2/2)\right]^{2/5}. \label{etf}
\end{equation}
One can see that in TF limit the energies of pure vortex-free
state ($\kappa=0$) and the state with the single vortex
($\kappa=1$) are equal, hence $\Omega_c$ found from relation
\begin{equation}
\frac{\Omega_c(\kappa)}{2\pi}=\frac{E(\kappa)-E(0)}{2\pi\hbar
N\kappa}, \label{omc}
\end{equation}
[see \cite{rmp} and Eq.~(\ref{lz})], at $\kappa=1$, vanishes in TF
limit. Thus, one should take the kinetic energy of the BEC cloud
into account. Here this is done by the numerical solution of
Eq.~(\ref{varpar}) for the two values $\kappa=0,1$. The result is
$\rho_0=2.19$ (2.64), $\zeta_0=9.28$ (10.66), and $\rho_1=1.61$
(1.90), $\zeta_1=9.06$ (10.53),  in the case of the experimental
conditions Eq.~(\ref{seta}), (\ref{setb}), respectively. One then
finds $\Omega^{\rm A}_c/2\pi=65.9$ Hz, and $\Omega^{\rm
B}_c/2\pi=36.1$ Hz, where upper indices refer to the two above
sets of the trap parameters. The smallness of these values as
compared to the experimentally measured  is due to smallness of
the energy difference between the pure BEC states with $\kappa=0$
and 1.

To reconcile the result of calculations with the measurements in
the present approach, one should have in mind that, in fact, the
density of atoms does not vanish in the central dip
\cite{vortexp2}. The authors of Ref.~\cite{vortexp2} propose three
possible reasons for this: (i) oscillations of the vortex
filament, (ii) the presence of non-condensed atoms, and (iii)
insufficient resolution of the imaging optics as compared to the
vortex core radius of the BEC cloud. Here we propose the fourth
possible reason and admit that the quantum state of the BEC cloud
after the stirring is the superposition of pure quantum states
with the angular momenta per particle $L_z/N=0$ and $\hbar$, that
is, admitting $\kappa\not=1$ in the wave function Eq.~(\ref{psi})
of final quantum state. The presence of BEC atoms in the state
with zero angular momentum explains in a natural way a nonzero
density in the central dip.

\begin{figure}
\centerline {\epsfysize=7in \epsfbox{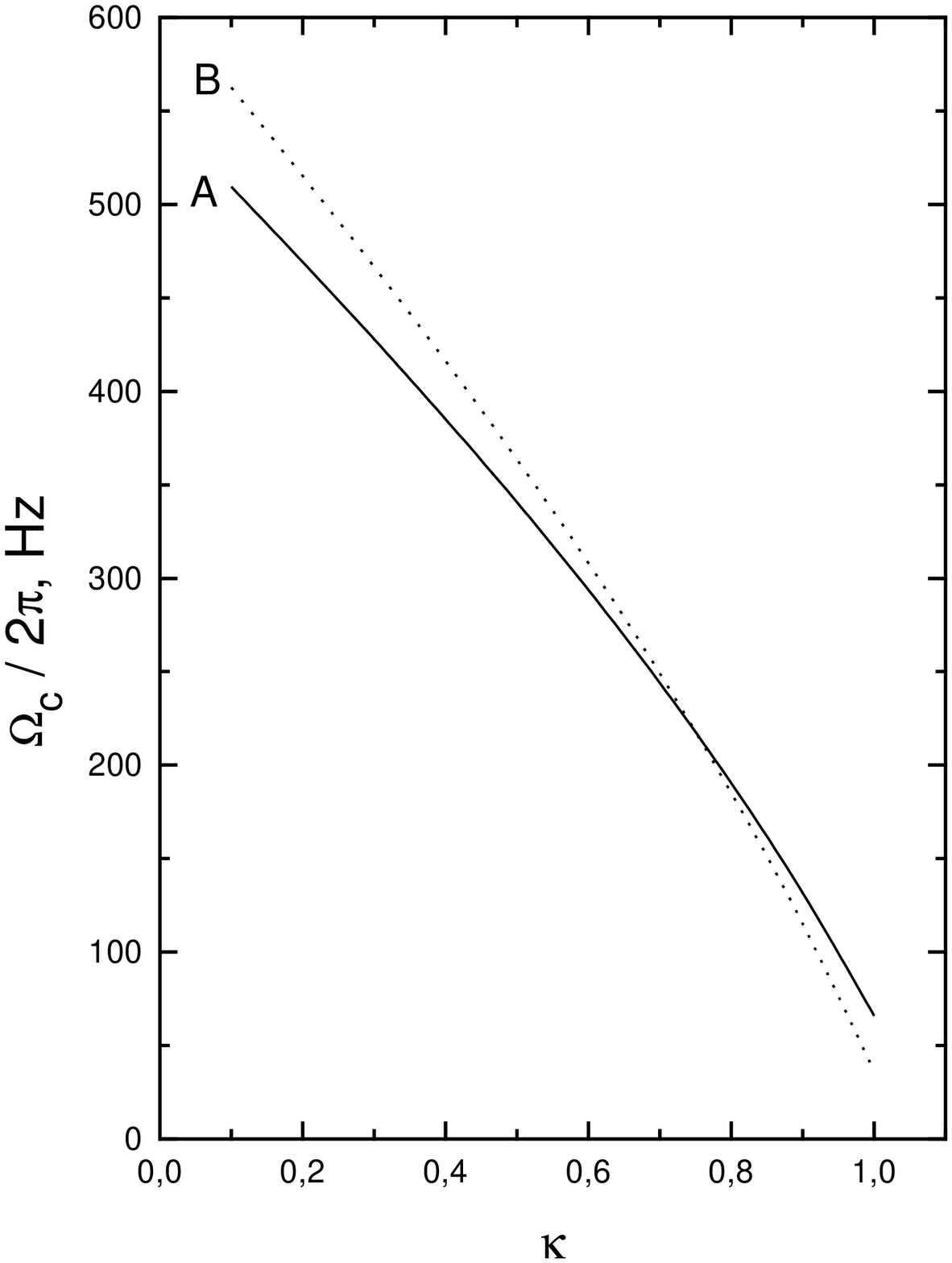}} \caption{The
dependence of the critical angular velocity of formation of the
quantum state of BEC with one vortex on the portion $\kappa$ of
atoms  in the state with the unit angular momentum per particle.
The curves labeled by A and B  correspond to the trap parameters
Eq.~(\ref{seta}) and (\ref{setb}), respectively. \label{fig1}}
\end{figure}

Solving Eq.~(\ref{varpar}) numerically, one can find the energy
dependence on $\kappa$ and calculate the critical angular velocity
from Eq.~(\ref{omc}). The result of this calculation is  shown in
Fig.~\ref{fig1}. Then fitting the calculated critical angular
velocity Eq.~(\ref{omc}) to the experimental values is possible if
the portion $\kappa$ of the number of BEC particles in the state
with the unit angular momentum per particle amounts to,
respectively,
\begin{equation}
\kappa^{\rm A}=0.87\mbox{, }\kappa^{\rm B}=0.90.\label{kapcal}
\end{equation}
The latter values, in view of Eq.~(\ref{lz}), give $\langle
L_Z\rangle/N=0.87\hbar$, respectively, $0.9\hbar$, and do not
contradict to the measured magnitude \cite{vortexp3} of the mean
angular momentum of the condensate. Eq.~(\ref{kapcal}) means that
the admixture of the atoms in the zero angular momentum state in
the case of the trap parameters Eq.~(\ref{seta}) [(\ref{setb})]
amounts to 0.13 (0.1), respectively. Notice that despite essential
difference in the experimental trap parameters Eqs.~(\ref{seta})
\cite{vortexp2} and (\ref{setb}) \cite{vortexp3}, the portion of
the number of atoms in the zero momentum state needed to explain
very different observed critical angular velocities in the final
state prepared after the stirring, turns out to be practically
same. To be more precise, the 30 $\%$ accuracy of the
determination of the number of atoms in the condensates reported
in Refs.~\cite{vortexp2,vortexp3} implies, as is explained earlier
in this paper, approximately $10\%$ uncertainty of the calculation
of energy per condensed atom which is translated to approximately
the same uncertainty of calculation of $\kappa$, while the central
values of the calculated $\kappa$ differ in the above experimental
conditions by $3 -4 \%$, which is well below their estimated
uncertainty.

Could such proposed feature of the BEC wave function as the
presence of the portion $1-\kappa$ of  zero angular momentum
condensed atoms be tested in experiments? Let us discuss this
issue. As is known \cite{vortexp2,vortexp3}, the presence of the
vortex is detected through the visualization of the images of BEC
cloud obtained after its ballistic expansion. One can obtain the
spatial distribution of the BEC atoms after this expansion with
the usual quantum mechanical method upon  finding the wave
function Eq.~(\ref{psi}) in the momentum space form, propagating
it forward in time freely, then finding its resulting coordinate
space form. The resulting spatial distribution appears to be
\begin{eqnarray}
\rho(r_\bot,\phi,z,t)&=&\frac{N}{\pi^{3/2}R^2_\kappa
z_\kappa}\left(1+\frac{\hbar^2t^2}{m^2z^4_\kappa}\right)^{-1/2}
\left(1+\frac{\hbar^2t^2}{m^2R^4_\kappa}\right)^{-1}
\exp\left[-\frac{r^2_\bot}{R^2_\kappa+\left(\frac{\hbar
t}{mR_\kappa}\right)^2}-\frac{z^2}{z^2_\kappa+\left(\frac{\hbar
t}{mz_\kappa}\right)^2}\right]\times\nonumber\\
&&\times\left[1-\kappa+\frac{\kappa
r^2_\bot/R^2_\kappa+2\sqrt{\kappa(1-\kappa)}r_\bot\left(\cos\phi+
\frac{\hbar
t}{mR^2_\kappa}\sin\phi\right)/R_\kappa}{1+\left(\frac{\hbar
t}{mR^2_\kappa}\right)^2}\right] . \label{rho}
\end{eqnarray}
One can see that the   dependence of the spatial distribution of
the BEC atoms on the azimuth angle $\phi$  is the signature of the
above admixture of the vortex free state. The dependence arises
from the interference term. But the contribution of the latter
becomes significant only after some time of duration of the
process of free expansion. Taking the estimates of the root mean
squared values of $r_\bot/R_\kappa$ and $r^2_\bot/R^2_\kappa$ from
the written spatial distribution function, one can find that the
axial asymmetric contribution $\propto\sin\phi$ becomes to
dominate after the time of flight
\begin{equation}
t>\tau\approx\frac{mR^2_\kappa}{\hbar}\sqrt{\frac{\kappa}{2(1-\kappa)}}.
\label{tof}
\end{equation}
Using the parameter sets Eqs.~(\ref{seta}), [(\ref{setb})], and
the results of numerical solutions of Eq.~(\ref{varpar}) together
with Eq.~(\ref{kapcal}), one can find from Eq.~(\ref{rho}) that
the interference term becomes to dominate after the free expansion
time exceeding $\tau=24$ ms (50 ms), respectively. Since the time
of free expansion reported in Refs.~\cite{vortexp2,vortexp3} is 27
ms, it is clear that the proposed feature of the final wave
function that could emerge after the stirring has  not enough time
to develop in the experiments \cite{vortexp2,vortexp3}. An
additional testable  feature of the proposed wave function of the
final state is the flattening  of the central vortex dip in the
density distribution due to the increasing relative contribution
of the zero angular momentum state $\propto 1-\kappa$ as time of
free expansion of the BEC cloud is increasing. It would be
interesting to enlarge (if possible) the time of ballistic
expansion of the BEC clouds after the stirring to see if the
dependence of their spatial distribution will acquire the angular
dependence $\propto\sin\phi$, and to study the relative weights of
the  components of condensed atoms with different angular momentum
at different times of their ballistic expansion.

\end{document}